\documentclass[aps,pra,showpacs,twocolumn]{revtex4-1}
\usepackage{amssymb}
\usepackage{amsmath}
\usepackage{graphicx}
\usepackage{epsfig}
\usepackage{subfigure}
\usepackage{appendix}

\begin{document}

\title{Coherent confinement and unidirectional dynamics of a wave packet induced by a non-Hermitian Su-Schrieffer-Heeger segment}
\author{K. L. Zhang}
\author{Z. Song}
\email{songtc@nankai.edu.cn}

\affiliation{School of Physics, Nankai University, Tianjin 300071, China}

\begin{abstract}

The competition between staggered imaginary potentials and lattice dimerization results in a unidirectional propagating coalescing state of a non-Hermitian Su-Schrieffer-Heeger (SSH) ring at an exceptional point. A segment of the SSH ring inherits the unidirectional feature and exhibits some intriguing dynamic behaviors when it is embedded in a Hermitian chain as a $\mathcal{PT}$-symmetric scattering center. Based on the Bethe ansatz scattering solution for the interface between Hermitian and non-Hermitian regions, we show that the SSH scattering center supports the following wave-packet dynamics. (i) A left incident wave packet is allowed perfect transmission, while a right incident wave packet stimulates an amplified reflecting wave train with the length proportional to the size of the SSH segment. Accordingly, a unidirectional invisibility of Bragg scatterers is observed when a multi-SSH segment is considered. (ii) A left incident wave packet can be well confined in a SSH segment if the right lead is removed. (iii) In addition, a wave packet is perfectly absorbed when its size closes to the length of the SSH segment. The underlying mechanism stems from a subtle property of the SSH segment, which supports two quasicoalescing wave packets but with opposite group velocities. Our findings are applicable to the schemes of realizing quantum state storage, quantum diode, and lasing device.
\end{abstract}

\maketitle

\section{Introduction}

\label{Introduction}

\begin{figure}[tbh]
\centering
\includegraphics[width=0.48\textwidth]{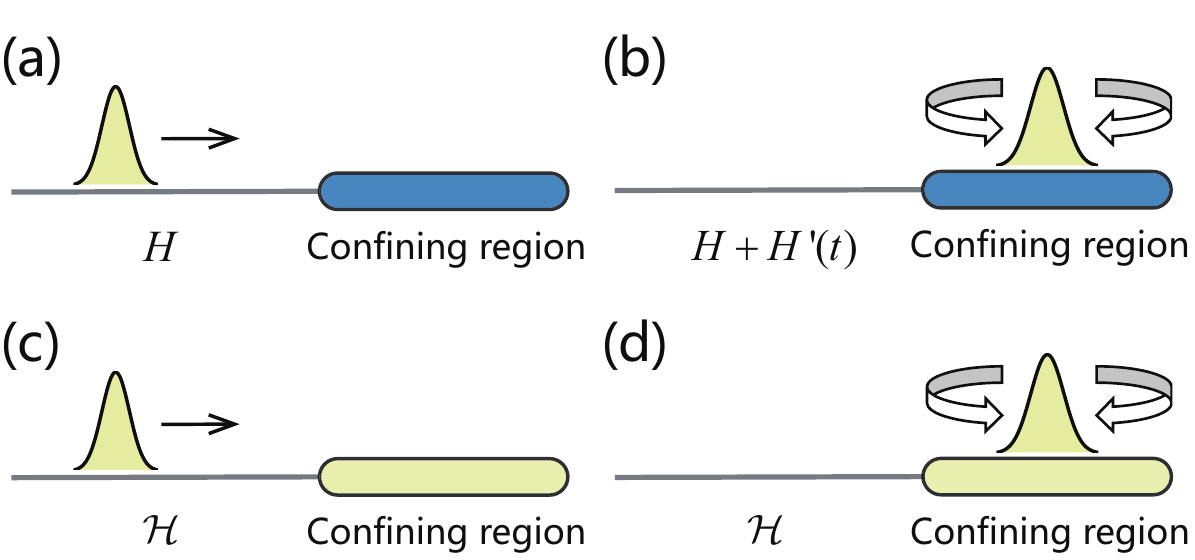}
\caption{Schematic illustration of the coherent confinement. (a) and (b) is
the time-dependent Hermitian system; (c) and (d) is the time-independent
non-Hermitian system.}
\label{Figure1}
\end{figure}

The introduction of complex potential in a Hamiltonian extends the frontier
of conventional quantum mechanics. Non-Hermitian systems make many things
possible including unidirectional propagation and anomalous transport \cite%
{Kulishov,LonghiOL,Lin,Regensburger,Eichelkraut,Feng,Peng,Chang}, invisible
defects \cite{LonghiPRA2010,Della,ZXZ}, coherent absorption \cite{Sun} and
self sustained emission \cite{Mostafazadeh,LonghiSUS,ZXZSUS,Longhi2015,LXQ},
loss-induced revival of lasing \cite{PengScience}, as well as laser-mode
selection \cite{FengScience,Hodaei}. Such kinds of novel phenomena can be
traced to the existence of exceptional point (EP). On the other hand,
implementing the tasks of quantum information processing via non-Hermitian
systems becomes an attractive topic. Recently, critical behavior of
non-Hermitian system has been employed to generate entangled states in a
dynamical process \cite{Tony1,Tony2, CLi2015, SLin2016}. It is well known
that a static Hamiltonian cannot realize a coherent confinement of a
wavepacket in the framework of Hermitian quantum mechanics, which is a\ key
ingredient for the storage of quantum information. It is always implemented
by time-dependent system \cite{MFYanik1, MFYanik2, SLonghiPRE, HWHQIP},
employing adiabatic or diabatic dynamical process. It seem impossible to
stop and confine a wavepacket via a natural time evolution under an
always-on Hermitian system. As illustrated in Figs. \ref{Figure1}(a) and
(c), a wave packet can be transferred from certain location to a confining
region by a natural time evolution. This process can be achieved by a
time-independent Hermitian Hamiltonian. In order to confine the wave packet,
a time-dependent term should be added (Fig. \ref{Figure1}(b)). A simplest
way is to switch off the tunneling channel at way out. However, it may be
done by a natural time evolution under a time-independent non-Hermitian
Hamiltonian (Fig. \ref{Figure1}(d)) since a non-Hermitian system exhibits
strong unidirectional dynamics \cite%
{LXQ,Longhi,JLSR,JLSR2,CLi17,JLNJP,JLPRL,ZXZ2019,ZKLPRB}. It is expected
that a wavepacket can perfectly come in the confining region without any
reflection, but do not leak out from the region. In addition, a compound
system that consists of non-Hermitian and Hermitian segments, such as a
sandwich structure, may possess some unrevealed features to be explored. 

In this paper, we consider whether it is possible to use time-independent
non-Hermitian systems to realize a coherent confinement of wavepacket via
natural time evolution. We investigate the scattering problem of the
interface between a uniform Hermitian and non-Hermitian Su-Schrieffer-Heeger
(SSH)-type semi-infinite chains via Bethe ansatz technique. It describes a
junction of two types of materials, Hermitian and non-Hermitian, which have
different properties. The essence is that the competition between staggered
imaginary potentials and lattice dimerization in a SSH model result in a
unidirectional propagating coalescing state at EP. And a segment of the SSH
ring inherits the unidirectional feature and exhibits some intriguing
dynamic behaviors when it is embedded in a Hermitian chain as a $\mathcal{PT}
$-symmetric scattering center (sandwich). Analytical and numerical results
indicate that a SSH scattering center supports the following wave packet
dynamics: (i) A left incident wave packet is allowed perfect transmission,
while a right incident wave packet stimulates an amplified reflecting wave
train with the length proportional to the size of the SSH segment. A
unidirectional invisibility of Bragg scatterers is observed when
multi-embedded-SSH-segment is considered. (ii) A left incident wave packet
can be well confined in SSH segment if the right lead is removed. (iii) In
addition, a wave packet is perfectly absorbed when its size closes to the
length of the SSH segment. The underlying mechanism relates to the peculiar
dynamics of wave packet at EP. Two wave packets with maximal opposite group
velocities can cancel each other out due to the fact that the main
components of two wave packets are the same coalescing state. In contrast,
two similar wave packets for Hermitian uniform chain are always orthogonal
in the context of Dirac inner product. The proposed schemes do not require
complicated structure, which is accessible in optical system. Our findings
are applicable to the schemes of realizing quantum state storage, quantum
diode and lasing device.

This paper is organized as follows. In Sec. \ref{Scattering at interface},
we present the model consisted of two semi-infinite chains and the
scattering solutions at the interface. In Sec. \ref{Non-Hermitian SSH
segment as a scattering center}, we discuss the dynamics of the system with
the non-Hermitian SSH segment as a scattering center. In Sec. \ref{Coherent
confinement and perfectly absorbed}, we demonstrate the coherent confinement
and perfectly absorbed dynamics of the semi-infinite system. In Sec. \ref%
{Summary}, we summarize our findings.

\section{Scattering at interface}

\label{Scattering at interface}

\begin{figure}[tbp]
\centering
\includegraphics[width=0.48\textwidth]{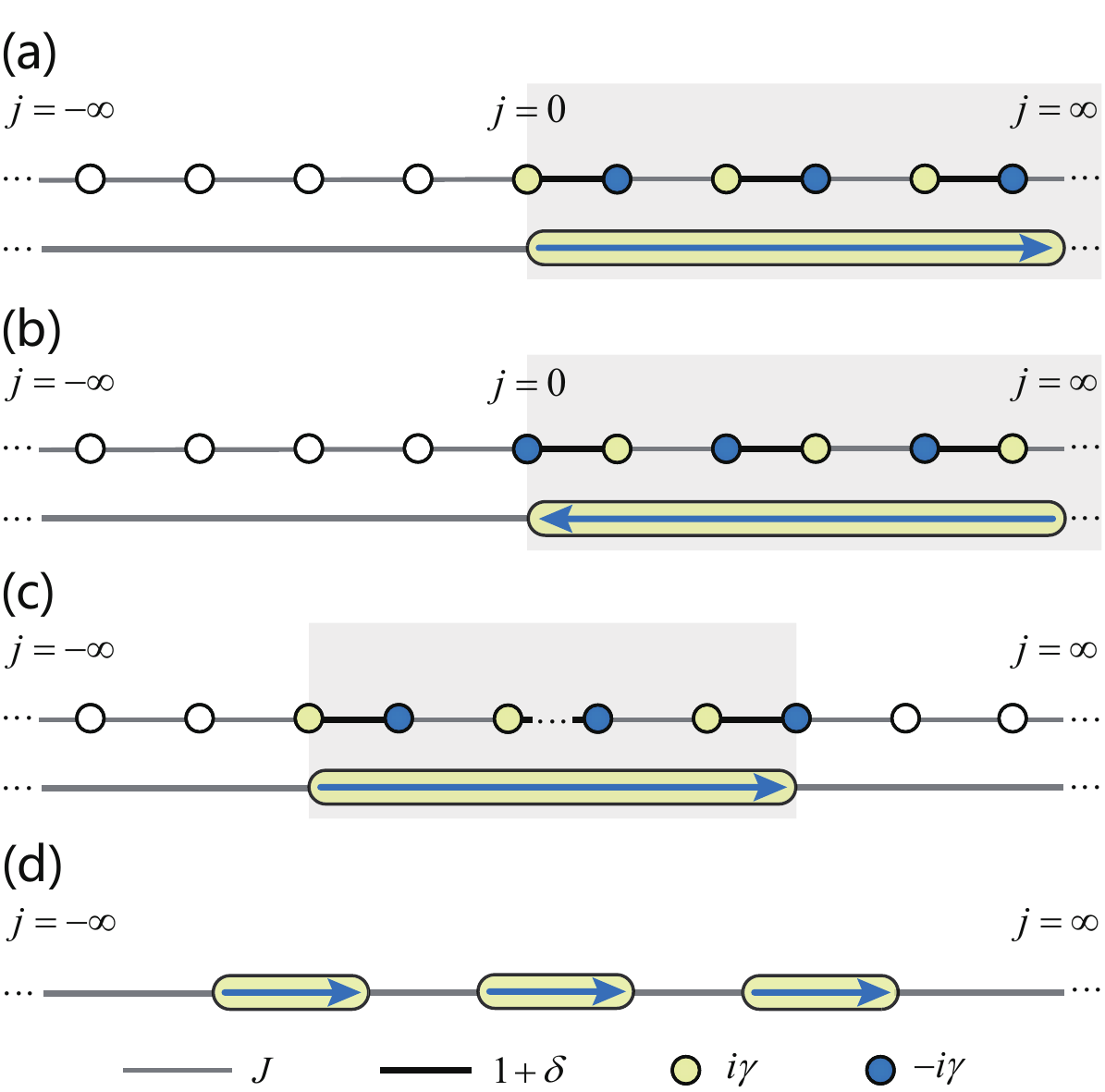}
\caption{Schematic illustration of the non-Hermitian infinite chain
representing by Eq. (\protect\ref{H}), which is consist of semi-infinite
subsystems: The uniform chain and the non-Hermitian SSH chain. (a) The
uniform chain is connected to the gain site of the non-Hermitian SSH chain.
(b) The uniform chain is connected to the loss site of the non-Hermitian SSH
chain. (c) A finite non-Hermitian SSH segment embedded in a uniform chain as
a $\mathcal{PT}$-symmetric scattering center. (d) A multi-layer sandwich
structure, in which many SSH segments are embedded in a uniform chain. The
gray line and blue arrow represent the uniform chain and non-Hermitian SSH
chain, respectively.}
\label{Figure2}
\end{figure}

One of the exclusive features of a non-Hermitian system is the
unidirectional dynamical behavior, which cannot occur in a Hermitian system
even the parity symmetry is broken. For instance, a non-Hermitian SSH chain,
the competition between staggered imaginary potentials and lattice
dimerization result in a unidirectional propagating coalescing state at EP 
\cite{ZKLPRA2,ZKLPRB}, while a linear field breaks the isotropy of a uniform
Hermitian chain in both directions, but\ can only lead to a Bloch
oscillation, exhibiting reversibility of wavepacket dynamics \cite%
{FBloch,CZener}. An interesting question is what happens when such two
systems are jointed together to form an infinite chain. Figures \ref{Figure2}%
(a) and (b) are the schematic illustrations of this configuration. In the
following, we will investigate the scattering solution for such an
interface, which is expected to shed light on the property of the interface
between a Hermitian and non-Hermitian system in general.

The corresponding Hamiltonian reads%
\begin{equation}
\mathcal{H}=H_{\mathrm{I}}+H_{\mathrm{II}},  \label{H}
\end{equation}%
which is consist of two semi-infinite chains. Here $H_{\mathrm{I}}$\
describes a Hermitian uniform semi-infinite chain%
\begin{equation}
H_{\mathrm{I}}=J\sum_{j=-\infty }^{-1}a_{j}^{\dag }a_{j+1}+\mathrm{H.c.},
\label{H1}
\end{equation}%
and $H_{\mathrm{II}}$ is a non-Hermitian semi-infinite SSH chain%
\begin{eqnarray}
H_{\mathrm{II}} &=&\sum_{l=0}^{\infty }[\left( 1+\delta \right) a_{2l}^{\dag
}a_{2l+1}+Ja_{2l+1}^{\dag }a_{2l+2}+\mathrm{H.c.}  \notag \\
&&+i\gamma (a_{2l}^{\dag }a_{2l}-a_{2l+1}^{\dag }a_{2l+1})],  \label{H2}
\end{eqnarray}%
where $a_{j}$\ and $a_{j}^{\dag }$\ are particle operators of fermion or
boson. The interface of the Hermitian and non-Hermitian regions is at $j=0$.
We notice that $\mathcal{H}$ breaks the time reversal symmetry and $\mathcal{%
H}(\gamma )=\mathcal{H}^{\ast }(-\gamma )=\mathcal{H}^{\dag }(-\gamma )$. In
this paper, we focus on the situation with the parameter satisfying $\gamma
=\pm \left( 1+\delta -J\right) $, which are shown in Figs. \ref{Figure2}(a)
and (b). Before we deal with the solution of $\mathcal{H}$, we briefly
review the features of solutions of $H_{\mathrm{I}}$\ and $H_{\mathrm{II}}$\
on an $N$-site ring lattice, which indicate the motivation of the above
parameter setting.

\begin{figure*}[tbh]
\centering
\includegraphics[width=1\textwidth]{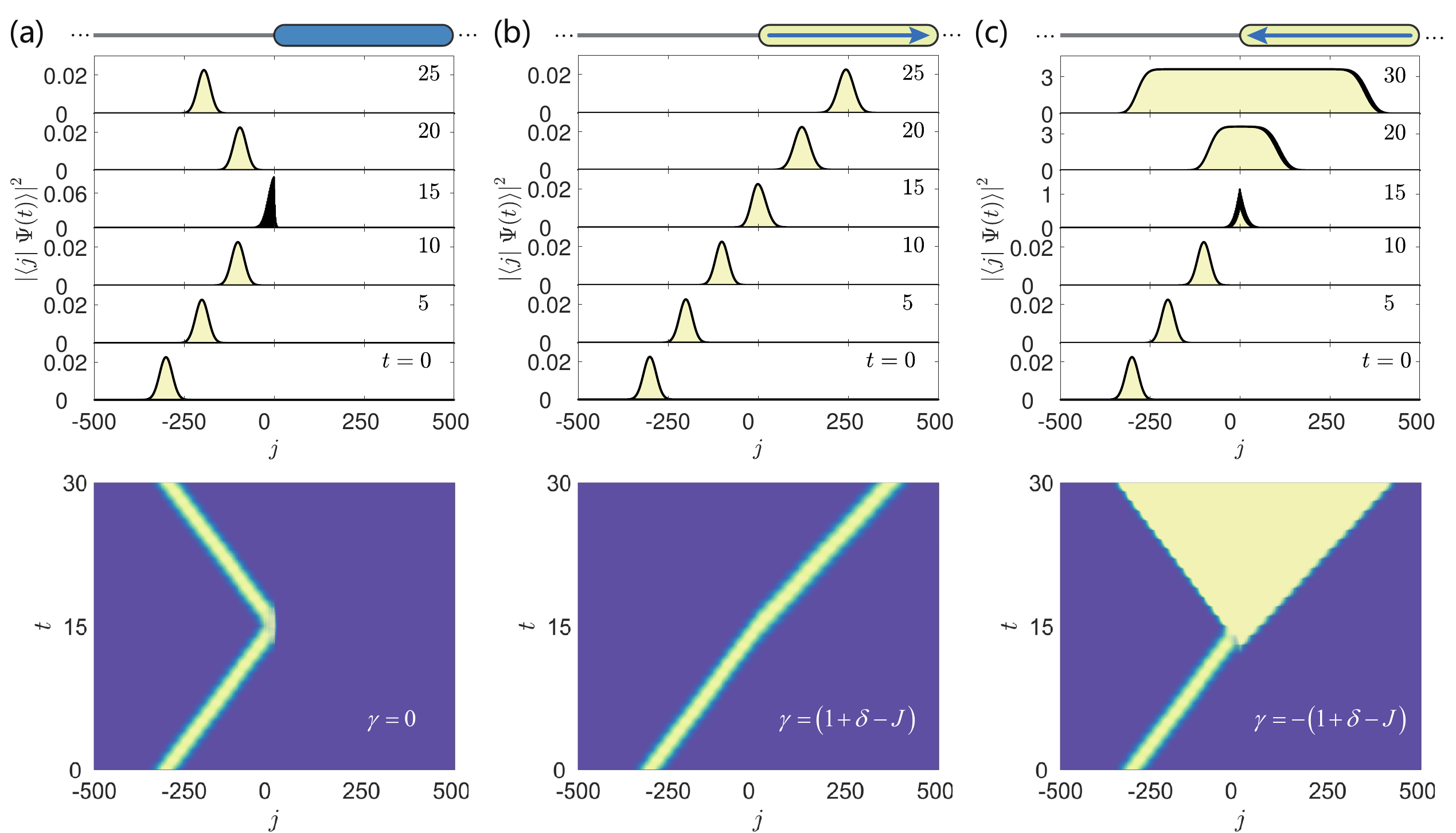}
\caption{Numerical simulations of the dynamics for three typical situations.
The system is consist of a uniform chain and a SSH chain. (a) Dynamics of
perfect reflection for the Hermitian lattice, which is consist of a uniform
chain and a Hermitian SSH chain. (b) Perfect transmission without refection
for the system in Fig. \protect\ref{Figure1}(a). (c) Amplified transmission
and reflection for the system in Fig. \protect\ref{Figure1}(b). The
parameters of the initial excitation are $\protect\alpha =0.04$, $N_{\mathrm{%
\ c}}=-300$ and $k_{\mathrm{c}}=-\protect\pi /2$. The parameters of the
system are $J=1$, $\protect\delta =0.5$ and $\protect\gamma =0$ for (a); $%
\protect\gamma =0.5$ for (b); $\protect\gamma =-0.5$ for (c). The time unit
is $10J^{-1}$. }
\label{Figure3}
\end{figure*}

For the uniform ring, $H_{\mathrm{I}}$ supports two degenerate zero modes in
the form of $\left\vert \psi _{\pm }\right\rangle
=N^{-1/2}\sum_{j=1}^{N}e^{\pm i\left( \pi /2\right) j}\left\vert
j\right\rangle $. Obviously, both directions are symmetric. The
corresponding ring system for non-Hermitian $H_{\mathrm{II}}$ is at EP when
takes $\gamma =\pm \left( 1+\delta -J\right) $. It supports a single
coalescing zero-energy eigenstate: $\left\vert \psi _{+}\right\rangle $ or $%
\left\vert \psi _{-}\right\rangle $, depending on the parameter $\gamma =\pm
\left( 1+\delta -J\right) $, which exhibits strongly unidirectionality.
Thus, it is expected that the infinite system $\mathcal{H}$ support a zero
energy scattering wave function $\sum_{j=-\infty }^{\infty }e^{-i\frac{\pi }{%
2}j}\left\vert j\right\rangle $ or $\sum_{j=-\infty }^{\infty }e^{i\frac{\pi 
}{2}j}\left\vert j\right\rangle $, corresponding to the parameter $\gamma
=\left( 1+\delta -J\right) $ or $\gamma =-\left( 1+\delta -J\right) $,
respectively. Furthermore, it has been shown that $H_{\mathrm{II}}$ still
has many intriguing features, such as simple harmonic oscillation and
position-independent lasing \cite{ZKLPRA1, ZKLPRA2}, even for the open
boundary condition. The underlying mechanism of these phenomena, that is
also the heart of this work, is the unidirectionality inherited from the SSH
ring at EP.

Now we turn to the combination of such two systems, by solving the
scattering problem of the interface between Hermitian and non-Hermitian
regions as the scattering center. In general, the ansatz wave function in
single-particle subspace has the form 
\begin{equation}
\left\vert \psi \right\rangle =\sum_{j=-\infty }^{\infty }f_{j}\left\vert
j\right\rangle ,
\end{equation}%
with 
\begin{equation}
f_{j}=\left\{ 
\begin{array}{cc}
\mathcal{I}e^{iKj}+\mathcal{O}e^{-iKj}, & j<0 \\ 
\mathcal{I}^{\mathrm{A}}e^{-ikj}+\mathcal{O}^{\mathrm{A}}e^{ikj}, & j=2l \\ 
\mathcal{I}^{\mathrm{B}}e^{-ikj}+\mathcal{O}^{\mathrm{B}}e^{ikj}, & j=2l+1%
\end{array}%
\right. .
\end{equation}%
For simplicity, we focus on the system with parameter $\gamma =\pm \left(
1+\delta -J\right) $ and the solutions with zero energy. The solution of
scattering wave function can be obtained by the Bethe ansatz method (see
Appendix A).

For the case with $\gamma =\left( 1+\delta -J\right) $, it is shown that the
scattering solution is%
\begin{equation}
\left\vert \psi \right\rangle =\sum_{j=-\infty }^{\infty }e^{-i\frac{\pi }{2}%
j}\left\vert j\right\rangle ,  \label{sol_a}
\end{equation}%
which is a plane wave,\ supporting the perfect transmission without
refection for the wave packet with momentum center $\pi /2$. It means that
the wave packet propagates with Dirac probability preservation and without
shape deformation. However, for $\gamma =-\left( 1+\delta -J\right) $, the
scattering solution is 
\begin{equation}
\left\vert \tilde{\psi}\right\rangle =\sum_{j=-\infty }^{\infty }e^{i\frac{%
\pi }{2}j}\left\vert j\right\rangle ,  \label{sol_b}
\end{equation}%
which can be obtained by directly applying the time-reversal operator on the
system and the solution. We list the two solutions explicitly for the use of
constructing the scattering solution for a non-Hermitian SSH segment as a
scattering center.

\begin{figure*}[tbh]
\centering
\includegraphics[width=1\textwidth]{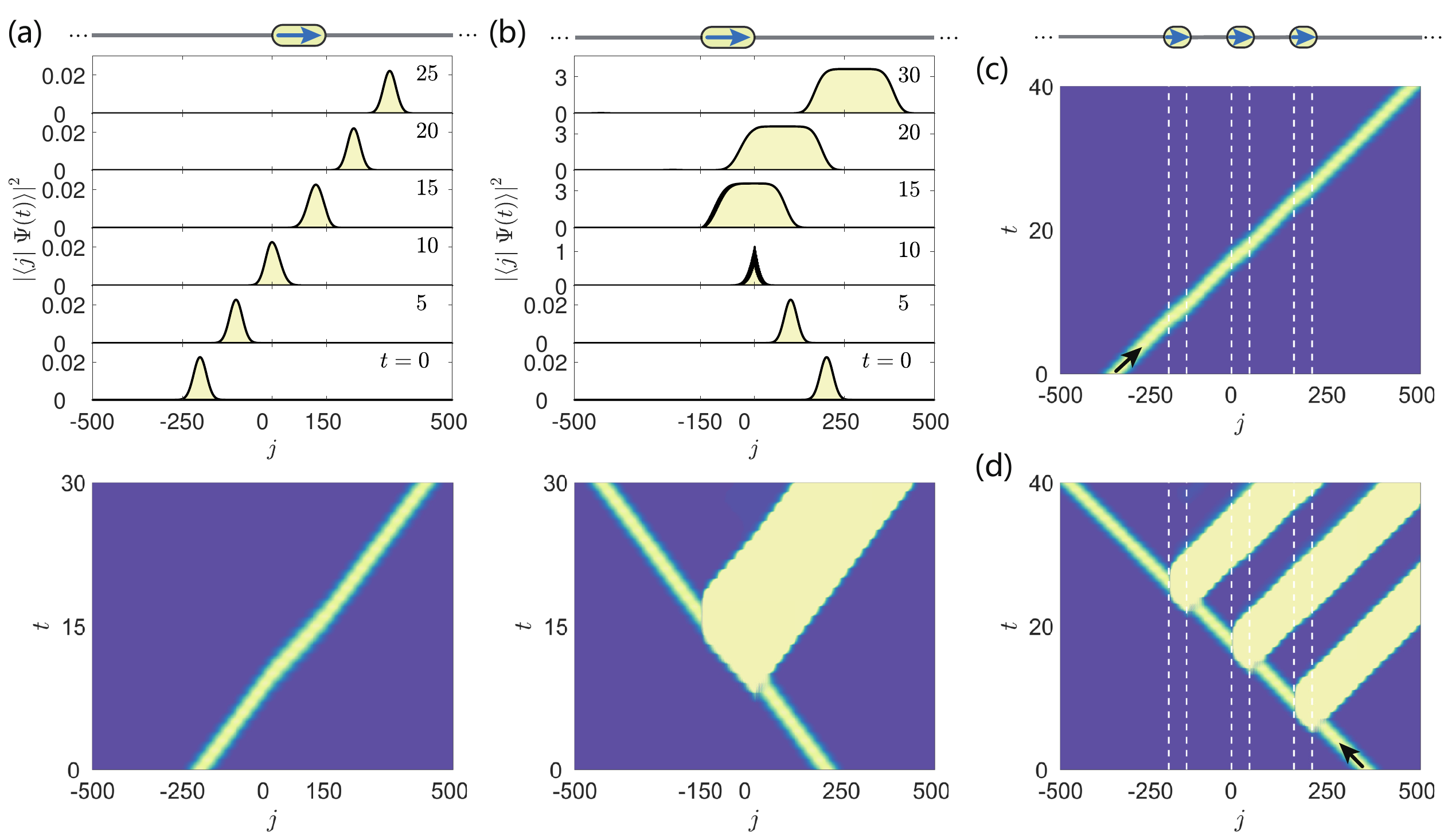}
\caption{Numerical simulations of the dynamics for the coupled system, which
is from by a uniform chain embedded with a non-Hermitian SSH segment. (a)
Perfect transmission without refection. (b) Amplified reflecting wave train.
The parameters of the initial excitation are $\protect\alpha=0.04$, $N_{ 
\mathrm{c}}=-200$ and $k_{\mathrm{c}}=-\protect\pi/2$ for (a); $\protect%
\alpha=0.04$, $N_{\mathrm{c}}=200$ and $k_{\mathrm{c}}=\protect\pi/2$ for
(b). The length of the segment is $\Delta=150$. (c) and (d) Numerical
results for the Bragg-stack-like scattering center with initial Gaussian
wave packet incoming from left and right of the system, respectively. The
length of three segments are all $\Delta=50$. Other parameters are $J=1$, $%
\protect\delta=0.5$ and $\protect\gamma=0.5$. The time unit is $10J^{-1}$.}
\label{Figure4}
\end{figure*}

The above two solutions indicate the unidirectionality of the system, which
is arisen from the non-Hermitian part of the system in Eq. (\ref{H2}), i.e.,
the competition between staggered imaginary potentials and lattice
dimerization. In general, a plane wave solution can be utilized to predict
the wave packet dynamics. To demonstrate this feature, we perform the
numerical simulation of the time evolution for an initial Gaussian wave
packet 
\begin{equation}
\left\vert \Psi \left( 0\right) \right\rangle =\Omega ^{-1/2}\sum_{j=-\infty
}^{\infty }e^{-\alpha ^{2}\left( j-N_{\mathrm{c}}\right) ^{2}/2}e^{ik_{%
\mathrm{c}}j}\left\vert j\right\rangle ,  \label{Gaussian}
\end{equation}%
where $N_{\mathrm{c}}$ is the wave packet center, $k_{\mathrm{c}}$ is the
central momentum, and $\Omega =\sqrt{\pi }/\alpha $\ is the normalization
factor. The evolved state obeys the equation $\left\vert \Psi \left(
t\right) \right\rangle =e^{-i\mathcal{H}t}\left\vert \Psi \left( 0\right)
\right\rangle $, which can be computed by exact diagonalization of a finite
system with sufficient large size. As comparison, we focus on three cases
with $\gamma =0$ and $\gamma =\pm \left( 1+\delta -J\right) $. The numerical
results for three typical situations are shown in Fig. \ref{Figure3}. For
the system in Fig. \ref{Figure2}(b), the scattering solution corresponding
to an incident plane wave with momentum $-\pi /2$ incoming from the left of
system can not be obtained through the scattering theory for plane waves.
The dynamics for perfect transmission without refection in Fig. \ref{Figure3}%
(b) accords with the previous analysis. In Fig. \ref{Figure3}(c), we can see
that for the system in Fig. \ref{Figure2}(b), an incident plane wave
incoming from the left of the system gives rise to amplified transmission
and reflection. So far, the mechanism of this behavior is unclear, but it is
not the focus of the present work. The dynamics of wavepacket for
non-Hermitian system in Figs. \ref{Figure3}(b) and (c) can not be seen in a
Hermitian system in Fig. \ref{Figure3}(a), which is consist of a uniform
chain and a Hermitian SSH chain by taking $\gamma =0$. In this Hermitian
system, the energy levels of the uniform chain does not match the one of the
Hermitian SSH chain, i.e., the plane wave mode with momentum $\pm \pi /2$ is
forbidden due to the energy gap of the Hermitian SSH chain. Thus the wave
packet cannot pass through the interface of two chains, exhibiting the
perfect reflection.

\begin{figure*}[tbh]
\centering
\includegraphics[width=1\textwidth]{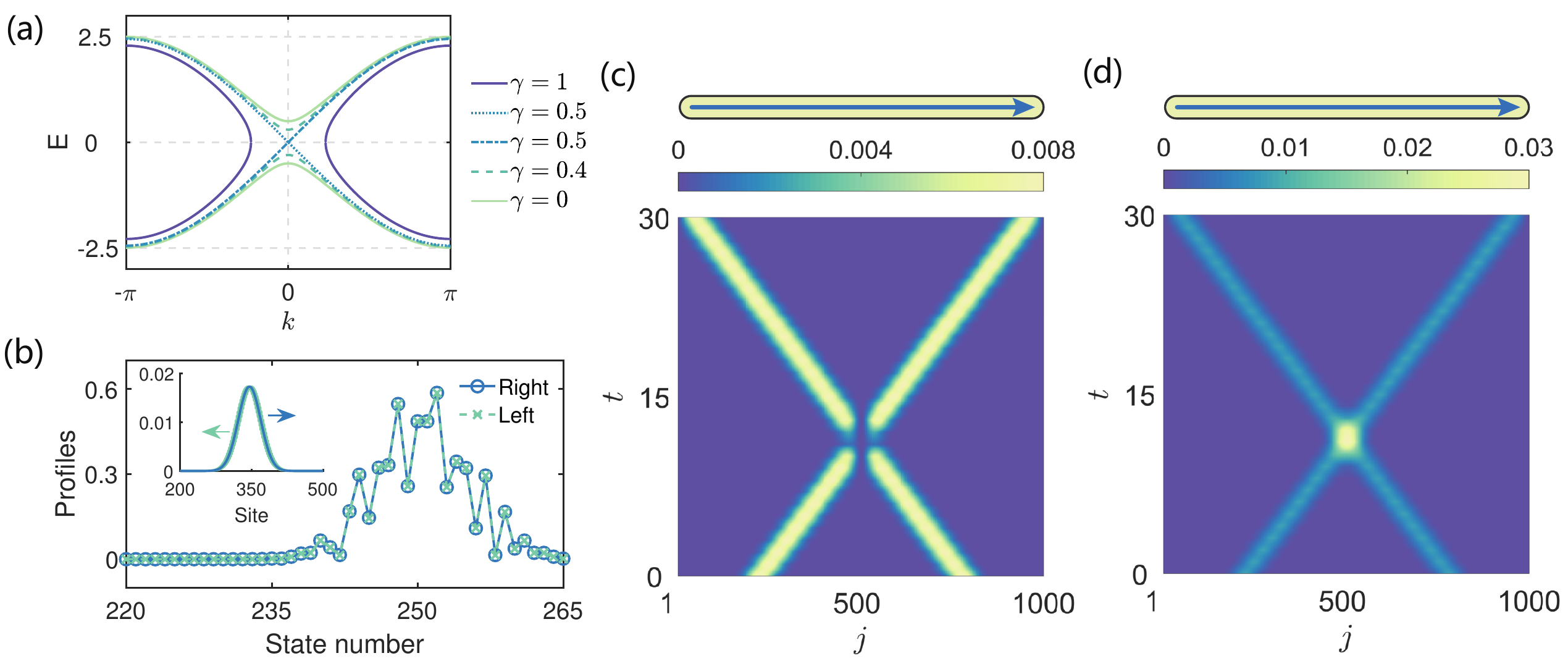}
\caption{(a) Energy dispersion of the non-Hermitian SSH ring in Eq. (\protect
\ref{dispersion}) for several typical value of $\protect\gamma$ when $J=1$
and $\protect\delta=0.5$. Here the energy is in unit of $J$. (b) Profiles of two wave packets ($\left\vert 
\protect\psi _{\mathrm{L}}\right\rangle$ and $\left\vert \protect\psi _{%
\mathrm{R}}\right\rangle$) with opposite group velocities in $k$ space and
real space (inset figure). Here the state number represents the number of
eigen states sorted by the energy from small to large. The size of the chain
system is $N=500$, and other parameters are $J=1$, $\protect\delta=0.5$ and $%
\protect\gamma=0.5$. (c) and (d) are the numerical simulations of wave
packet dynamics in the non-Hermitian SSH chain with the initial states: (c) $%
\left(\left\vert \protect\psi _{\mathrm{L}}\right\rangle+\left\vert \protect%
\psi _{\mathrm{R}}\right\rangle\right)/\protect\sqrt{2}$, and (d) $%
\left(\left\vert \protect\psi _{\mathrm{L}}\right\rangle-\left\vert \protect%
\psi _{\mathrm{R}}\right\rangle\right)/\protect\sqrt{2}$. The size of the
system is $N=1000$, and other parameters are $J=1$, $\protect\delta=0.5$ and 
$\protect\gamma=0.5$. The time unit is $10J^{-1}$.}
\label{Figure5}
\end{figure*}

\section{Non-Hermitian SSH-segment scattering center}

\label{Non-Hermitian SSH segment as a scattering center}

In Sec. \ref{Scattering at interface}, we have discussed the scattering
features at the interface of Hermitian uniform chain and non-Hermitian SSH
chain in specific case. Both the Hermitian and non-Hermitian systems are
semi-infinite. It reveals the unidirectional feature of non-Hermitian SSH
model. In this section, we consider the case with a finite non-Hermitian SSH
segment embedded in a uniform chain as a $\mathcal{PT}$-symmetric scattering
center. Schematics of such a system is given in Fig. \ref{Figure2}(c). A
solution of such a sandwich can be obtained by the combination of Eqs. (\ref%
{sol_a}) and (\ref{sol_b}), when the number of sites of the non-Hermitian
SSH segment is even $\Delta $, which ensures the balance of gain and loss.
Furthermore, such a resonant transmission solution can be extended to a
multi-layer sandwich structure, in which many SSH segments are embedded in a
uniform chain (see Fig. \ref{Figure2}(d)).

This feature can be demonstrated via wave packet dynamics. We perform the
numerical simulation for the time evolution of Gaussian wave packets with
two initial situations. The numerical results for the wave packet incoming
from the left and the right of the system are shown in Figs. \ref{Figure4}%
(a) and (b), respectively. In Fig. \ref{Figure4}(a), we can see that the
left-incident wave packet pass through the non-Hermitian SSH segment without
refection, and the probability is conserved. This accords with the
prediction from Bethe ansatz solution. Actually, in certain non-Hermitian systems, the dynamics of probability preservation can be realized when the initial states are structured properly, as shown in the previous works \cite{ZKLPRA1,JLPRA2011,HWH2012,ZKLPRB2}. In Appendix C, we show that when the initial states are structured properly in a solvable non-Hermitian SSH ring, the Dirac probability conserve. Thus, together with the Bethe ansatz solution and the analysis in Appendix C, it is presumable that when certain type of wavepacket pass through the interface and evolve in a non-Hermitian SSH segment, the dynamics of probability preservation can be observed. In Fig. \ref{Figure4}(b), we can see
that the right-incident wave packet can trigger the full transmission
associating with amplified reflecting wave train. The width of the amplified
wave train $h$ is about twice as the size of the non-Hermitian SSH segment: $%
h\approx 2\Delta =300$. So for, this observation cannot be explained by the
present scattering solution. We leave this issue for the future
consideration.

The behavior of perfect transmission in Fig. \ref{Figure4}(a) can be well
understood. In Sec. \ref{Scattering at interface}, we have discussed the
scattering solutions for the systems in Figs. \ref{Figure2}(a) and (b). And
the sandwich structure in upper panel of Fig. \ref{Figure4}(a) can be
regarded as the combination of two system in Figs. \ref{Figure2}(a) and (b),
thus the scattering solutions in Eqs. (\ref{sol_a}) and (\ref{sol_b}) can
explain the dynamics of perfect transmission in Fig. \ref{Figure4}(a).
Accordingly, a perfect transmission in\ multi-layer sandwich structure can
be predicted by repeating the process in Fig. \ref{Figure4}(a). In Figs. \ref%
{Figure4}(c) and (d), numerical results for a Bragg-stack-like scattering
center are plotted, which is similar to the experimental observation for
unidirectional invisibility of $\mathcal{PT}$-symmetric Bragg scatterers 
\cite{Regensburger}.

\section{Coherent confinement and perfect absorption}

\label{Coherent confinement and perfectly absorbed} 
\begin{figure*}[tbh]
\centering
\includegraphics[width=1\textwidth]{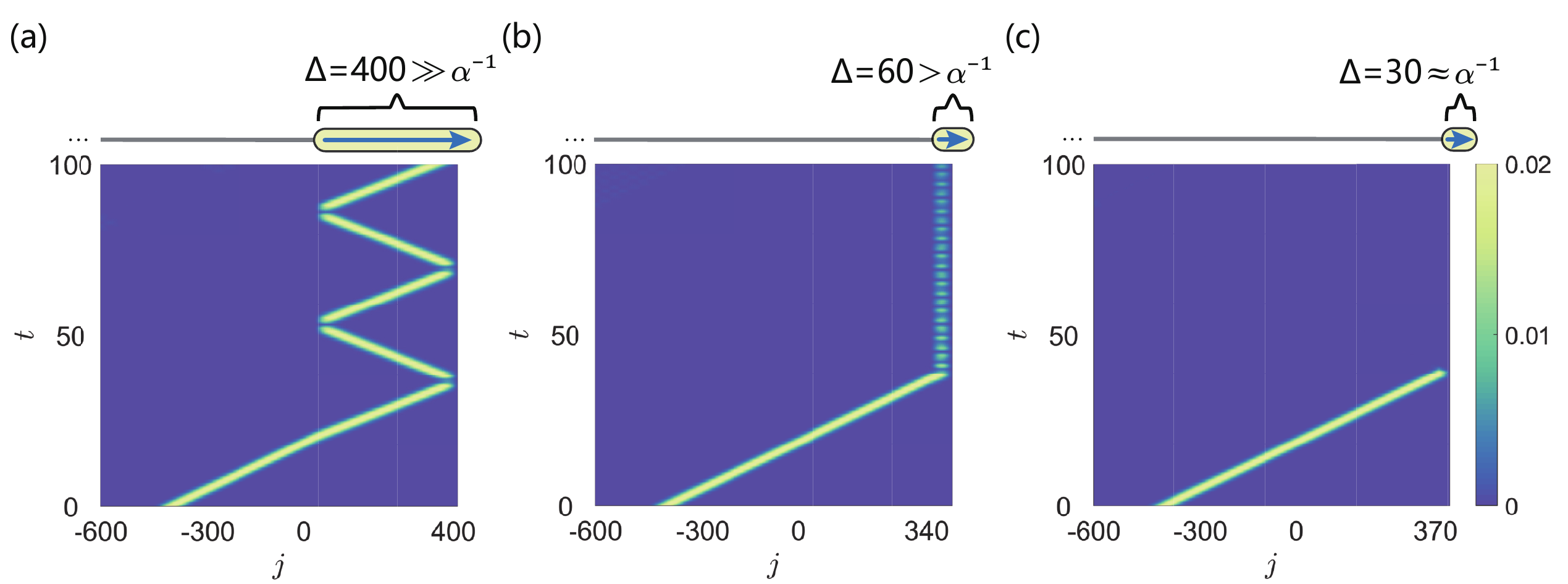}
\caption{Numerical simulations of the dynamics for the semi-infinite system,
which is consist of a uniform semi-infinite chain and a non-Hermitian SSH
segment. (a) Coherent confinement when the width of the wave packet is much
smaller than the size of the segment. (b) When the width of the wave packet
approach to size of the segment, the confined probability decrease. (c)
Perfectly absorbed when the width of the wave packet approximate equal to
the size of the segment. The parameters of the initial excitation are $%
\protect\alpha =0.04$, $N_{\mathrm{c}}=-400$ and $k_{\mathrm{c}}=-\protect%
\pi /2$; and for the system are $J=1$, $\protect\delta =0.5$ and $\protect%
\gamma =0.5$. The time unit is $10J^{-1}$. }
\label{Figure6}
\end{figure*}
Before proceeding with this section we would like to reveal a peculiar
feature of wave packet dynamics in a non-Hermitian SSH system at EP, which
is crucial for the mechanism of the phenomena that will be presented in this
section. We start with the dispersion relation of the SSH model, which
determines the group velocity of a wave packet. The dispersion relation is
expressed as%
\begin{equation}
\varepsilon _{k}^{\pm }=\pm \sqrt{J^{2}+\left( 1+\delta \right)
^{2}-2J\left( 1+\delta \right) \cos k-\gamma ^{2}},  \label{dispersion}
\end{equation}%
which is plotted in Fig. \ref{Figure5}(a) for several typical values of $%
\gamma $. We are interested in the case with $\gamma =0.5$, at which a
coalescing state appears at $k=0$. In the vicinity of zero $k$, the
dispersion relation is two crossing lines with the crossing point at EP. One
can construct two wave packets ($\left\vert \psi _{\mathrm{L}}\right\rangle $
and $\left\vert \psi _{\mathrm{R}}\right\rangle $ as the form in Eq. (\ref%
{Gaussian})) with zero energy, but with opposite group velocities, which
correspond to the slopes of the linear dispersion lines. According to the
analysis in Appendix D, there exist two quasi-coalescing sets of energy levels
near the zero point. On the other hand, one can do the same thing for a
Hermitian uniform chain (or the SSH model with $\gamma =\delta =0 $).
However, two situations are different: (i) The zero point of the dispersion
relation for the later is degenerate point; (ii) The two wave packets for
the later are always orthogonal, while two wave packets in the non-Hermitian
system have nonzero overlap in the context of Dirac inner product. In Fig. %
\ref{Figure5}(b) we plot the profiles of two wave packets in $k$ space and
real space. We get the conclusion that when two such wave packets meet
together, the evolved state is a single wave packet with an additional
overall factor $1+e^{i\varphi }$. They can cancel out each other for $%
\varphi =\pi $, or double the amplitude for $\varphi =0$. To demonstrate
this results, numerical simulation is performed and the results are plotted
in Figs. \ref{Figure5}(c) and (d). The motivation of this study stems from
the following paradox. (On the one hand,) an infinite SSH chain supports a $%
\pi /2$ plane wave, while a zero-energy solution is forbidden for a
semi-infinite SSH chain (see Appendix B). The above analysis about the
zero-energy wave packets indicates that the physics of the forbidden mode is
the long-period cancelation of two wave packets of large width.

One may be curious what happens for the wave packet dynamics when one of the
Hermitian leads is removed and the system become semi-infinite. To answer
this question, we apply the above analysis on this semi-infinite sandwich
system. For the Hermitian lead, there are two zero energy channels of wave
packets with opposite group velocities. In contrast, there is only one zero
energy channel of wave packet, although it allows a wave packet propagating
with two opposite group velocities. A coalescing wave packet in the SSH
segment cannot escape via the lead through which it comes. Thus once a wave
packet comes into the SSH segment as perfect transmission, it bounces up and
down within the segment. Furthermore, the Dirac probability within the
segment is periodic due to the interference effect between incident and
reflected parts of the wave packet at boundaries.

To verify the above predictions, we perform the numerical simulation for the
evolution of a Gaussian wave packet in the form of Eq. (\ref{Gaussian}). The
numerical results are shown in Fig. \ref{Figure6}. In Fig. \ref{Figure6}(a),
the size of the segment is $\Delta =400$. We can see that when the width of
the wave packet is much smaller than the size of the non-Hermitian SSH
segment, the wave packet is confined in the segment. Figure \ref{Figure6}(b)
show that when the width of the wave packet approaches to the size of the
segment $\Delta =60$, the confined probability is reduced. In Fig. \ref%
{Figure6}(c), the size of the segment is $\Delta =30$. We can see the when
the width of the wave packet approximate equal to the size of the segment $%
\Delta \approx \alpha ^{-1}=25$, the wave packet is perfectly absorbed.
These results accords with our predictions.

\section{Summary}

\label{Summary} In summary, we have studied wave packet dynamics in the
compound system which consists of a non-Hermitian SSH segment and a
Hermitian uniform chain. Both systems support left and right propagating
wave packets with zero energy. However, two wave packets in the former are
quasi-coalescing, matching only one of two channels (left or right) in the
Hermitian chain. This leads to unidirectional dynamics at the interface of
the two subsystems. A wave packet in an SSH segment embedded in a uniform
chain (a sandwich structure) is either perfectly transmitted or totally
reflected at the interface, depending on the direction of the SSH segment.
In addition, two quasi-coalescing wave packets interfere with each other
constructively or destructively when they meet together. These result in
peculiar phenomena which are exclusive in non-Hermitian systems: For a
sandwich structure, an incident wave packet is either allowed perfect
transmission or stimulates an amplified reflecting wave train with the
length proportional to the size of the SSH segment. In addition, for a
Hermitian-non-Hermitian junction system, an incident wave packet can be well
confined in or perfectly absorbed by the SSH segment, depending on the
length of the SSH segment. The model setup contains only staggered tunneling
strength and imaginary potentials, which has relative simple geometrical
structure and non-Hermitian components. It should be experimentally
accessible in optical system \cite{Regensburger,SWeimann,HZhao,MPan}. An
inspired manifestation is the unidirectional invisibility of Bragg
scatterers for the simulation of multi-embedded-SSH-segment system. Our
findings are applicable to the schemes of realizing quantum state storage,
quantum diode and lasing device.

\section*{Appendix}

\setcounter{equation}{0} \renewcommand{\theequation}{A\arabic{equation}} %
\setcounter{figure}{0} \renewcommand{\thefigure}{A\arabic{figure}} %
\renewcommand{\thesubsection}{\Alph{subsection}}

In this Appendix, we present the derivations of (A) the scattering solutions
of the Hamiltonian in Eq. (\ref{H}); (B) the zero energy solution for
semi-infinite SSH chain; (C) and (D) the analysis for the solution of SSH
ring, introducing a concept of quasi-coalescing energy level sets.

\subsection{The scattering solutions of $\mathcal{H}$}

The Bethe ansatz wave function of a scattering state $\left\vert \psi
\right\rangle $ takes the form 
\begin{equation}
\left\vert \psi \right\rangle =\sum_{j=-\infty }^{\infty }f_{j}\left\vert
j\right\rangle .
\end{equation}%
with 
\begin{equation}
f_{j}=\left\{ 
\begin{array}{cc}
\mathcal{I}e^{iKj}+\mathcal{O}e^{-iKj}, & j<0 \\ 
\mathcal{I}^{\mathrm{A}}e^{-ikj}+\mathcal{O}^{\mathrm{A}}e^{ikj}, & j=2l \\ 
\mathcal{I}^{\mathrm{B}}e^{-ikj}+\mathcal{O}^{\mathrm{B}}e^{ikj}, & j=2l+1%
\end{array}%
\right. .
\end{equation}%
Here $l=0,1,2,...$ . The Schr\"{o}dinger equation $\mathcal{H}\left\vert
\psi \right\rangle =E\left\vert \psi \right\rangle $ gives the expression of
energy 
\begin{eqnarray}
E &=&2J\cos K \\
&=&\pm \sqrt{J^{2}-\gamma ^{2}+\left( 1+\delta \right) ^{2}+2J\left(
1+\delta \right) \cos \left( 2k\right) },  \notag
\end{eqnarray}%
and the equations for $\mathcal{I}$, $\mathcal{O}$, $\mathcal{I}^{\mathrm{A/B%
}}$, and $\mathcal{O}^{\mathrm{A/B}}$ 
\begin{eqnarray}
-\nu _{-K}\mathcal{O}+J\mathcal{O}^{\mathrm{A}} &=&\nu _{K}\mathcal{I}-J%
\mathcal{I}^{\mathrm{A}},  \notag \\
Je^{iK}\mathcal{O}-\mu _{-k}\mathcal{O}^{\mathrm{A}} &=&-Je^{-iK}\mathcal{I}%
+\mu _{k}\mathcal{I}^{\mathrm{A}},  \notag \\
\mathcal{I}^{\mathrm{B}} &=&\lambda _{k}\mathcal{I}^{\mathrm{A}},  \notag \\
\mathcal{O}^{\mathrm{B}} &=&\lambda _{-k}\mathcal{O}^{\mathrm{A}},
\end{eqnarray}%
in which, 
\begin{eqnarray}
\mu _{k} &=&\left( E-i\gamma \right) -\left( 1+\delta \right) \lambda
_{k}e^{-ik},  \notag \\
\nu _{K} &=&Ee^{-iK}-Je^{-2iK},  \notag \\
\lambda _{k} &=&\frac{Je^{-ik}+\left( 1+\delta \right) e^{ik}}{E+i\gamma }.
\end{eqnarray}
When $\gamma =\left( 1+\delta -J\right) $, the energy has the simple form 
\begin{equation}
E=2J\cos K=2\sqrt{J\left( 1+\delta \right) }\cos k.
\end{equation}
We consider the solutions with zero energy. Equation $E=0$ gives four
solutions for $K$ and $k$, that is $\left( K,k\right)=$ $\left( -\pi /2,-\pi
/2\right)$ ,$\left( \pi /2,\pi /2\right)$, $\left( \pi /2,-\pi /2\right)$, $%
\left( -\pi /2,\pi /2\right)$. However, it can be check that they correspond
to the same wave function 
\begin{equation}
\left\vert \psi \right\rangle =\sum_{j=-\infty }^{\infty }e^{-i\frac{\pi }{2}%
j}\left\vert j\right\rangle ,
\end{equation}

The Hamiltonian fulfill $\mathcal{H}^{\dag }=\mathcal{H}^{\ast },$ thus\ the
eigenstate with zero energy for $\mathcal{H}^{\dag }$ is 
\begin{equation}
\left\vert \tilde{\psi}\right\rangle =\left\vert \psi \right\rangle ^{\ast
}=\sum_{j=-\infty }^{\infty }e^{i\frac{\pi }{2}j}\left\vert j\right\rangle ,
\end{equation}

\begin{figure}[tbp]
	\centering
	\includegraphics[width=0.5\textwidth]{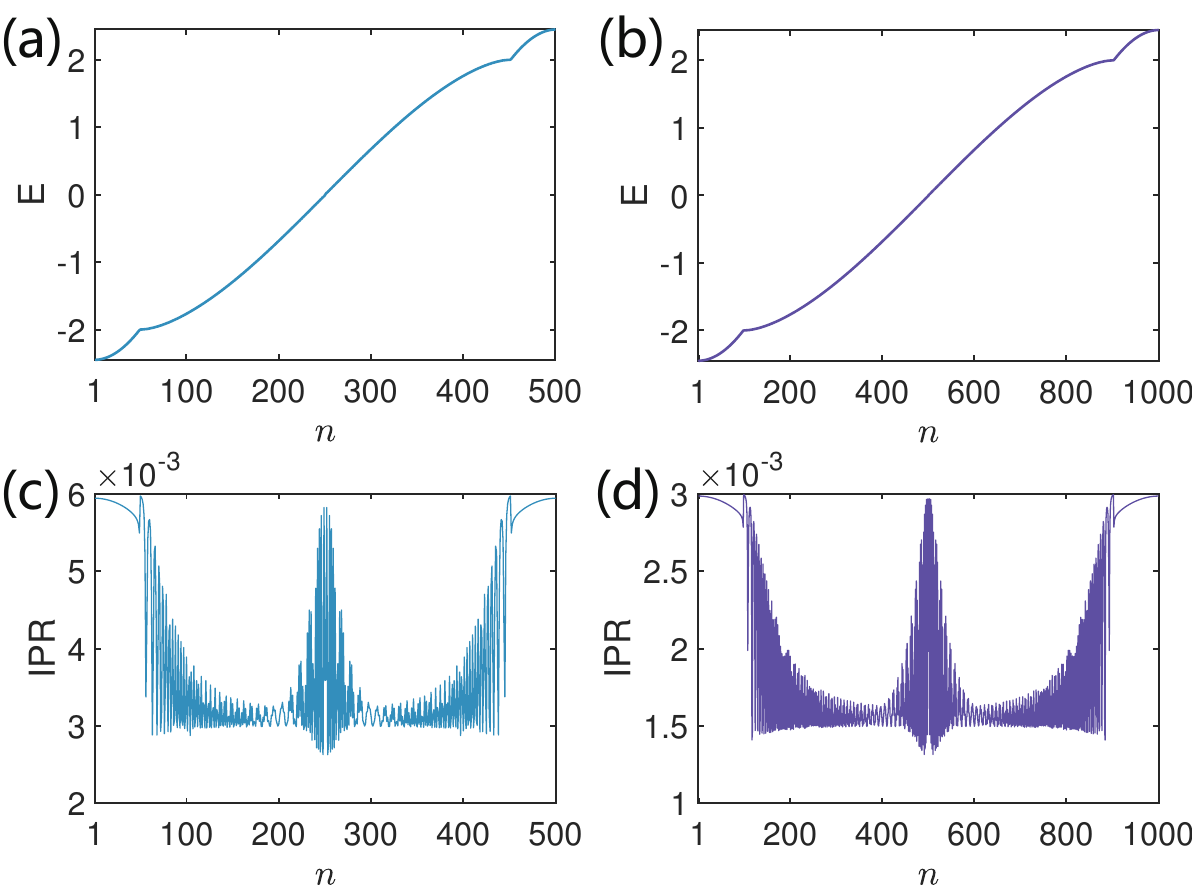}
	\caption{(a) and (b) are the energy spectra for the finite system obtained by numerical diagonalization. (c) and (d) are the IPR corresponding to (a) and (b), respectively. The size of system is $250+250$ for (a) and (c); and is $500+500$ for (b) and (d). Other parameters are $J=1$, $\protect\delta=0.5$ and $\protect\gamma=0.5$. Here the energy is in unit of $J$.}
	\label{FigureA}
\end{figure}

For the present model, it is tough to get the complete solutions analytically. Here we provide the complete solution for the finite system consisted of two subsystems (the uniform chain and the non-Hermitian SSH chain) by performing numerically exact diagonalization to clarify that the energy spectrum is real and all the eigenstates are not bound states. We introduce the inverse participation ratio (IPR), which is defined as $\mathrm{IPR}\left(n\right) =\sum_{j}\left\vert \left\langle j\right. \left\vert \psi _{n}\right\rangle \right\vert ^{4}$, with $n$ denotes the state number and $j$ denotes the lattice site, and the eigenstates $\left\vert \psi _{n}\right\rangle$ are Dirac normalized. In Fig. \ref{FigureA}, we plot the energy spectrum and the corresponding IPR with different lattice size. We observe that all the eigenstates of the finite system has real eigen energy. The IPR are finite and decay when the size of the system increase, which indicates that all the eigen states are not bound states for a system with large lattice size.

\subsection{Zero energy solution of the semi-infinite SSH chain}

\begin{figure}[tbp]
\centering
\includegraphics[width=0.4\textwidth]{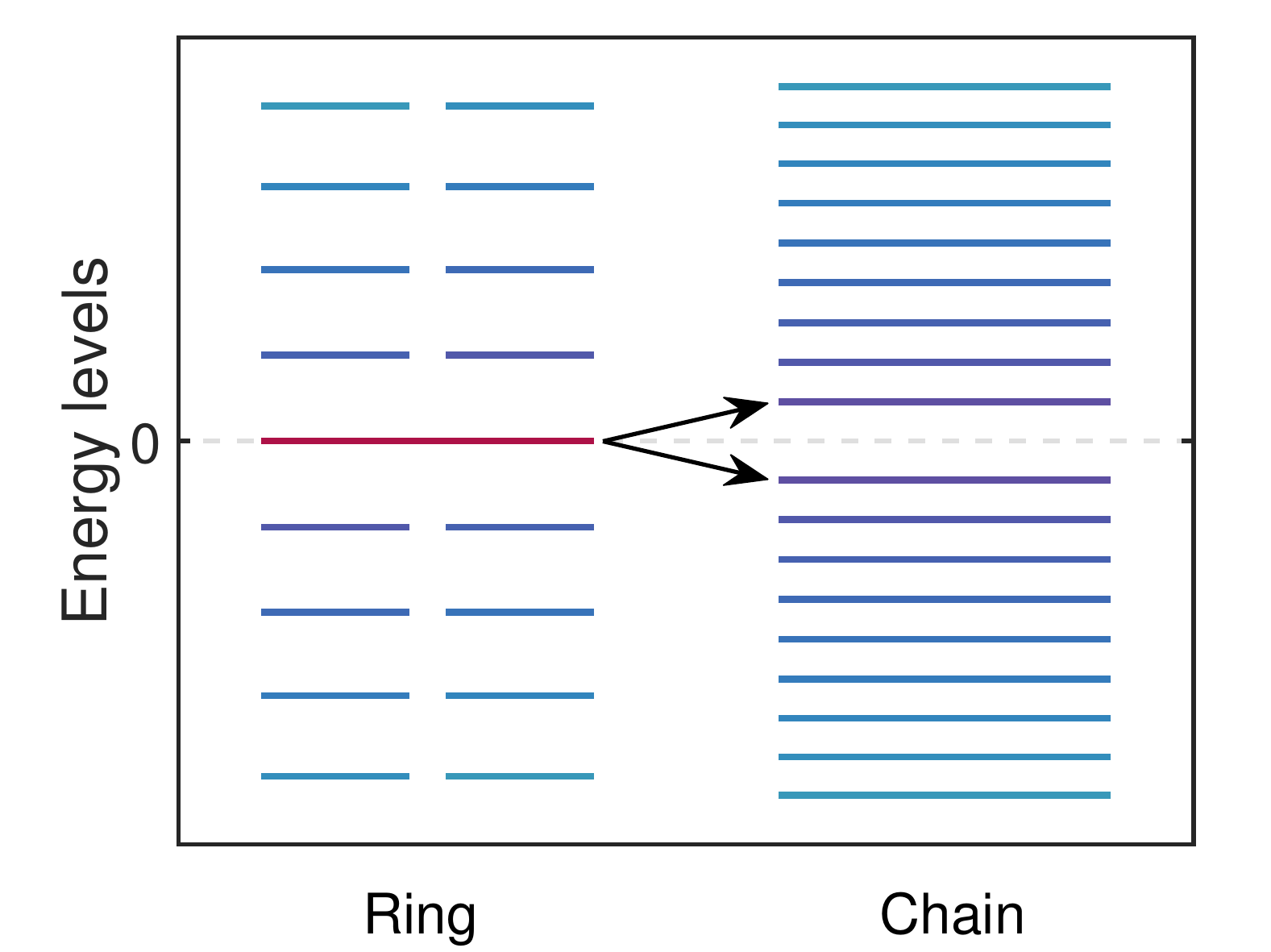}
\caption{Schematic illustration of the energy levels of the non-Hermitian
SSH model at EP with periodic and open boundary (by breaking one of the weak
hopping term) conditions, respectively. For a ring system, zero-level
coalescence is split into two levels when the open boundary condition is
imposed.}
\label{FigureA1}
\end{figure}

The ansatz wave function for the the semi-infinite SSH chain in Eq. (\ref{H2}%
) takes the form 
\begin{equation}
\left\vert \psi _{k}\right\rangle =\sum_{j=0}^{\infty }f_{j}\left\vert
j\right\rangle ,  \label{psik}
\end{equation}%
with 
\begin{equation}
f_{j}=\left\{ 
\begin{array}{cc}
\mathcal{I}^{\mathrm{A}}e^{-ikj}+\mathcal{O}^{\mathrm{A}}e^{ikj}, & j=2l \\ 
\mathcal{I}^{\mathrm{B}}e^{-ikj}+\mathcal{O}^{\mathrm{B}}e^{ikj}, & j=2l+1%
\end{array}%
\right. ,  \label{fjk}
\end{equation}%
The Schr\"{o}dinger equation $H_{\mathrm{II}}\left\vert \psi
_{k}\right\rangle =E_{k}\left\vert \psi _{k}\right\rangle $ gives 
\begin{equation}
\left( 1+\delta \right) \left( \mathcal{I}^{\mathrm{B}}e^{-ik}+\mathcal{O}^{%
\mathrm{B}}e^{ik}\right) +\left( i\gamma -E_{k}\right) \left( \mathcal{I}^{%
\mathrm{A}}+\mathcal{O}^{\mathrm{A}}\right) =0,  \label{j0}
\end{equation}%
and 
\begin{eqnarray}
\left[ i\gamma -E_{k}\right] \mathcal{I}^{\mathrm{A}}+\left[ \left( 1+\delta
\right) e^{-ik}+Je^{ik}\right] \mathcal{I}^{\mathrm{B}} &=&0,  \notag \\
\left[ Je^{-ik}+\left( 1+\delta \right) e^{ik}\right] \mathcal{I}^{\mathrm{A}%
}+\left[ -i\gamma -E_{k}\right] \mathcal{I}^{\mathrm{B}} &=&0,  \notag \\
\left[ i\gamma -E_{k}\right] \mathcal{O}^{\mathrm{A}}+\left[ \left( 1+\delta
\right) e^{ik}+Je^{-ik}\right] \mathcal{O}^{\mathrm{B}} &=&0,  \notag \\
\left[ Je^{ik}+\left( 1+\delta \right) e^{-ik}\right] \mathcal{O}^{\mathrm{A}%
}+\left[ -i\gamma -E_{k}\right] \mathcal{O}^{\mathrm{B}} &=&0.  \label{jn}
\end{eqnarray}%
Conditions for the existence of nonzero solutions are 
\begin{equation}
\left\vert 
\begin{array}{cc}
i\gamma -E_{k} & \left( 1+\delta \right) e^{-ik}+Je^{ik} \\ 
Je^{-ik}+\left( 1+\delta \right) e^{ik} & -i\gamma -E_{k}%
\end{array}%
\right\vert =0,
\end{equation}%
and 
\begin{equation}
\left\vert 
\begin{array}{cc}
i\gamma -E_{k} & \left( 1+\delta \right) e^{ik}+Je^{-ik} \\ 
Je^{ik}+\left( 1+\delta \right) e^{-ik} & -i\gamma -E_{k}%
\end{array}%
\right\vert =0,
\end{equation}%
which both give 
\begin{equation}
E_{k}=\pm \sqrt{J^{2}-\gamma ^{2}+(1+\delta )^{2}+2J(1+\delta )\cos (2k)}.
\label{Ek}
\end{equation}%
On the other hand, from Eqs. (\ref{j0}) and (\ref{jn}), we obtain 
\begin{eqnarray}
\mathcal{I}^{\mathrm{B}} &=&\frac{(E_{k}-i\gamma )e^{ik}}{1+\delta +Je^{2ik}}%
,  \notag \\
\mathcal{O}^{\mathrm{A}} &=&-\frac{\left[ (1+\delta )e^{2ik}+J\right] e^{2ik}%
}{1+\delta +Je^{2ik}},  \notag \\
\mathcal{O}^{\mathrm{B}} &=&-\frac{(E_{k}-i\gamma )e^{3ik}}{1+\delta
+Je^{2ik}}.  \label{IO}
\end{eqnarray}%
Here we set $\mathcal{I}^{\mathrm{A}}=1.$

We are looking for the solution with zero energy under the parameter $\gamma
=\pm \left( 1+\delta -J\right) $. Eq. (\ref{Ek}) is reduced to $E_{k}=2\sqrt{%
J(1+\delta )}\cos k=0$, which requires $k=\pm \pi /2.$ Substitute $\gamma
=\pm \left( 1+\delta -J\right) $ and $k=\pm \pi /2$ into Eqs. (\ref{fjk})
and (\ref{IO}), we have 
\begin{equation}
f_{j}=\left\{ 
\begin{array}{cc}
\pm \left( e^{-i\frac{\pi }{2}j}-e^{i\frac{\pi }{2}j}\right) , & j=2l \\ 
\pm \left( e^{-i\frac{\pi }{2}j}+e^{i\frac{\pi }{2}j}\right) , & j=2l+1%
\end{array}%
\right. .
\end{equation}%
$f_{j}$ is zero because $l=0,1,2,...,$ which means that the zero energy
solution is forbidden. This can be seen from the spectrum of the SSH chain
in Ref. \cite{ZKLPRA1, ZKLPRA2}. We demonstrate this point in Fig. \ref%
{FigureA1} by comparison two energy level structures of SSH model with open
and periodic boundary conditions. We see that the coalescing level with zero
energy in SSH ring disappears when the open boundary condition is imposed.
It accords with the vanishing zero energy solution for semi-infinite SSH
chain since the solution for an open chain can be obtained by matching the
solutions from two semi-infinite chains.

\subsection{Hermitian dynamics in a SSH ring}

For the purpose of explaining the Hermitian dynamics in the current
non-Hermitian system, we consider a solvable non-Hermitian SSH ring with the
Hamiltonian 
\begin{eqnarray}
H_{\mathrm{Ring}} &=&\sum_{l=1}^{N}[\left( 1+\delta \right) a_{2l-1}^{\dag
}a_{2l}+Ja_{2l}^{\dag }a_{2l+1}+\mathrm{H.c.}  \notag \\
&&+i\gamma (a_{2l-1}^{\dag }a_{2l-1}-a_{2l}^{\dag }a_{2l})].
\label{SSH_ring}
\end{eqnarray}%
In the $\mathcal{PT}$-symmetric region which support real eigen energy, the
Hamiltonian can be diagonalized as 
\begin{equation}
H_{\mathrm{Ring}}=\sum_{k}\left( \varepsilon _{k}^{+}\bar{\alpha}%
_{k,+}\alpha _{k,+}+\varepsilon _{k}^{-}\bar{\alpha}_{k,-}\alpha
_{k,-}\right) ,
\end{equation}%
with the eigen energy 
\begin{eqnarray}
\varepsilon _{k}^{\pm } &=&\pm \sqrt{g_{k}g_{k}^{\ast }-\gamma ^{2}}  \notag
\\
&=&\pm \sqrt{J^{2}+\left( 1+\delta \right) ^{2}-2J\left( 1+\delta \right)
\cos k-\gamma ^{2}},
\end{eqnarray}%
by taking the linear transformation 
\begin{eqnarray}
\bar{\alpha}_{k,\pm } &=&\frac{1}{\sqrt{2}}\left( \pm e^{i\theta _{k}^{\pm }}%
\sqrt{\frac{g_{k}}{g_{k}^{\ast }}}a_{k,1}^{\dag }+a_{k,0}^{\dag }\right) , \\
\alpha _{k,\pm } &=&\frac{\sqrt{2}}{e^{i\theta _{k}^{+}}+e^{i\theta _{k}^{-}}%
}\left( \sqrt{\frac{g_{k}^{\ast }}{g_{k}}}a_{k,1}+e^{-i\theta _{k}^{\pm
}}a_{k,0}\right) ,
\end{eqnarray}%
where 
\begin{eqnarray}
a_{k,\lambda } &=&N^{-1/2}\sum_{l=1}^{N}e^{i(k+\pi )l}a_{2l-\lambda
},(\lambda =1,0),  \notag \\
e^{i\theta _{k}^{\pm }} &=&\frac{\pm \sqrt{g_{k}g_{k}^{\ast }-\gamma ^{2}}%
+i\gamma }{\pm \sqrt{g_{k}g_{k}^{\ast }}},  \notag \\
g_{k} &=&\left( 1+\delta \right) -Je^{ik}.
\end{eqnarray}%
It can be checked that operators $\bar{\alpha}_{k,\pm }$ satisfy the
quasicanonical commutation relations 
\begin{eqnarray}
\left[ \bar{\alpha}_{k,+}^{\dag },\bar{\alpha}_{k^{\prime },+}\right] _{\pm
} &=&\left[ \bar{\alpha}_{k,-}^{\dag },\bar{\alpha}_{k^{\prime },-}\right]
_{\pm }=\delta _{kk^{\prime }},  \notag \\
\left[ \bar{\alpha}_{k,+}^{\dag },\bar{\alpha}_{k^{\prime },-}\right] _{\pm
} &=&2ie^{-i\theta _{k}^{+}}\sin \theta _{k}^{+}\delta _{kk^{\prime }}.
\label{comm}
\end{eqnarray}%
Here $\left[ .,.\right] _{\pm }$ denotes the the commutator and
anticommutator.

Now we consider the time evolution of an arbitrary initial state 
\begin{equation}
\left\vert \psi \left( 0\right) \right\rangle =\sum_{k}\left( C_{k,+}\bar{%
\alpha}_{k,+}+C_{k,-}\bar{\alpha}_{k,-}\right) \left\vert \mathrm{vac}%
\right\rangle .
\end{equation}
At instant $t$, we have 
\begin{eqnarray}
\left\vert \psi \left( t\right) \right\rangle &=&e^{-iHt}\left\vert \psi
\left( 0\right) \right\rangle  \\
&=&\sum_{k}\left( e^{-i\varepsilon _{k}^{+}t}C_{k,+}\bar{\alpha}%
_{k,+}+e^{-i\varepsilon _{k}^{-}t}C_{k,-}\bar{\alpha}_{k,-}\right)
\left\vert \mathrm{vac}\right\rangle .\notag
\end{eqnarray}
Using the quasicanonical commutation relations of Eqs. (\ref{comm}), the
Dirac probability can be written as 
\begin{eqnarray}
P\left( t\right) &=&\left\langle \psi \left( t\right) \right. \left\vert
\psi \left( t\right) \right\rangle   \\
&=&\sum_{k}\left( \left\vert C_{k,+}\right\vert ^{2}+\left\vert
C_{k,-}\right\vert ^{2}\right)  \notag \\
&&+4\sum_{k}\mathrm{Re}\left( -ie^{i\theta _{k}^{+}}e^{-2i\varepsilon
_{k}^{+}t}C_{k,+}C_{k,-}^{\ast }\right) \sin \theta _{k}^{+}.\notag
\end{eqnarray}

It indicate that when the initial states are structured properly: $%
C_{k,+}C_{k,-}^{\ast }\rightarrow 0$ (the initial state does not involve
components of $\bar{\alpha}_{k,+}$ and $\bar{\alpha}_{k,-}$ simultaneously),
the Dirac probability conserve.

\subsection{Quasi-coalescing levels in a SSH ring}

\begin{figure}[tbh]
	\centering
	\includegraphics[width=0.4\textwidth]{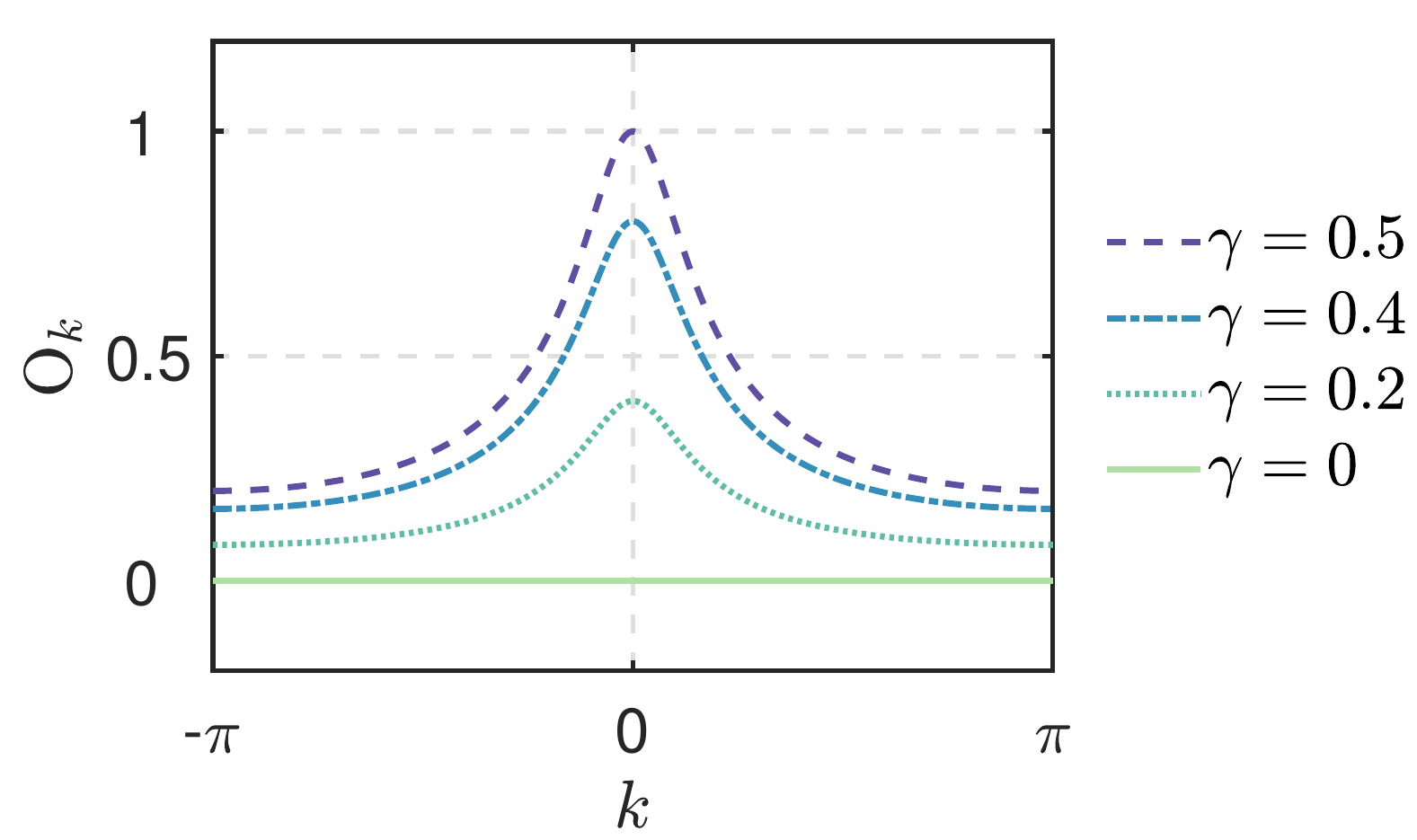}
	\caption{The overlap between the corresponding wave functions of two bands
		of the non-Hermitian SSH ring for several typical $\protect\gamma$. Other
		parameters are $J=1$ and $\protect\delta=0.5$.}
	\label{FigureA2}
\end{figure}

We consider a non-Hermitian SSH ring with the Hamiltonian of Eq. (\ref{SSH_ring}).
According to the analysis in Appendix C, the eigenstates of two bands $\varepsilon
_{k}^{\pm }$ can be obtained 
\begin{eqnarray}
\left\vert \psi _{k}^{\pm }\right\rangle  &=&\bar{\alpha}_{k,\pm }\left\vert 
\mathrm{vac}\right\rangle  \notag\\
&=&\frac{1}{\sqrt{2}}\left( \frac{\varepsilon _{k}^{\pm }+i\gamma }{1+\delta
-Je^{-ik}}a_{k,1}^{\dag }+a_{k,0}^{\dag }\right) \left\vert \mathrm{vac}%
\right\rangle .
\end{eqnarray}
Here the wave functions are Dirac normalized. We are interested in the overlap
between the corresponding wave functions of two bands, which is defined as%
\begin{equation}
\mathrm{O}_{k}=\left\vert \langle \psi _{k}^{+}\left\vert \psi
_{k}^{-}\right\rangle \right\vert .
\end{equation}%
Direct derivation shows that 
\begin{equation}
\mathrm{O}_{k}=\frac{\left\vert \gamma \right\vert }{\sqrt{J^{2}+\left(
1+\delta \right) ^{2}-2J\left( 1+\delta \right) \cos k}}.  \label{Ok}
\end{equation}%
At EP with $\gamma =\left( 1+\delta -J\right) $, we note that $\mathrm{O}%
_{k=0}=1$\ indicating the coalescence of state. We also find that $\mathrm{O}%
_{k}\approx 1$\ for small $k$,\ especially in the case of large $\gamma $,
which corresponds to strong dimerization limit. Actually, taking $\gamma
=\left( 1+\delta -J\right) $ and $\cos k\approx 1-k^{2}/2$, we have%
\begin{equation}
\mathrm{O}_{k}\approx \frac{1}{\sqrt{1+J\left( 1+\delta \right) k^{2}/\gamma
^{2}}},
\end{equation}%
which approaches to $1$ for large $\gamma $. We plot Eq. (\ref{Ok}) for
several typical $\gamma $\ in Fig. \ref{FigureA2} to demonstrate this point.
We conclude that there exist two sets of energy levels near the zero point,
which are quasi-coalescing. This is directly related to the vanishing zero energy
solution in SSH chain we discussed in the Appendix B.

\acknowledgments This work was supported by National Natural Science
Foundation of China (under Grant No. 11874225).

\end{document}